\documentclass[sigconf]{acmart}
\AtBeginDocument{%
  }

\usepackage{tcolorbox}
\usepackage{fancyvrb}
\usepackage{tikz}
\usepackage{todonotes}

\copyrightyear{2025}
\acmYear{2025}
\setcopyright{rightsretained}
\acmConference[SIGIR '25]{Proceedings of the 48th International ACM SIGIR
Conference on Research and Development in Information Retrieval}{July 13--18,
2025}{Padua, Italy}
\acmBooktitle{Proceedings of the 48th International ACM SIGIR Conference on
Research and Development in Information Retrieval (SIGIR '25), July 13--18,
2025, Padua, Italy}\acmDOI{10.1145/3726302.3730147}
\acmISBN{979-8-4007-1592-1/2025/07}

\begin{document}

\title{Artifact Sharing for Information Retrieval Research}

\author{Sean MacAvaney}
\email{sean.macavaney@glasgow.ac.uk}
\orcid{0000-0002-8914-2659}
\affiliation{%
  \institution{University of Glasgow}
  \city{Glasgow}
  \country{United Kingdom}
}

\renewcommand{\shortauthors}{MacAvaney}

\begin{abstract}
Sharing artifacts---such as trained models, pre-built indexes, and the code to use them---aids in reproducibility efforts by allowing researchers to validate intermediate steps and improves the sustainability of research by allowing multiple groups to build off one another's prior computational work. Although there are \textit{de facto} consensuses on how to share research code (through a git repository linked to from publications) and trained models (via HuggingFace Hub), there is no consensus for other types of artifacts, such as built indexes. Given the practical utility of using shared indexes, researchers have resorted to self-hosting these resources or performing \textit{ad hoc} file transfers upon request, ultimately limiting the artifacts' discoverability and reuse. This demonstration introduces a flexible and interoperable way to share artifacts for Information Retrieval research, improving both their accessibility and usability.

\vspace{0.4em}
\hspace{1.4em}\includegraphics[width=1.25em,height=1.25em]{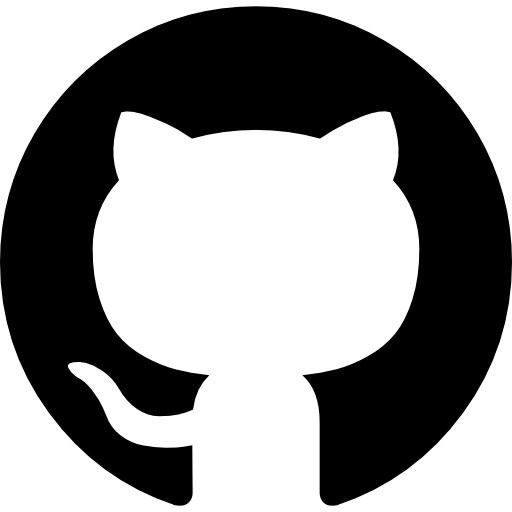}\hspace{.3em}
\parbox[c]{\columnwidth}
{
    \vspace{-.55em}
    \href{https://github.com/seanmacavaney/artifacts-demo}{\nolinkurl{https://github.com/seanmacavaney/artifacts-demo}}
}
\vspace{-1.2em}
\end{abstract}

\ccsdesc[500]{Information systems~Information retrieval}

\keywords{Reproduciblity, Artifact Reuse, Green IR}

\maketitle

\section{Introduction}

A variety of artifacts are produced throughout the course of research in Information Retrieval, including trained models, built indexes, caches, and the code to do it all. A few \textit{de facto} standard approaches have emerged for distributing some of these kinds of artifacts. For instance, code is typically shared via git repositories linked in the research paper, and trained models are typically shared on the HuggingFace Hub. However, there is no such standard for other types of artifacts---especially built indexes. This demonstration presents a new system that unifies access to data-oriented artifacts (such as built indexes and caches), while maintaining the flexibility required to conduct research in this fast-paced field.

\begin{figure}
\centering
\includegraphics[width=0.8\linewidth]{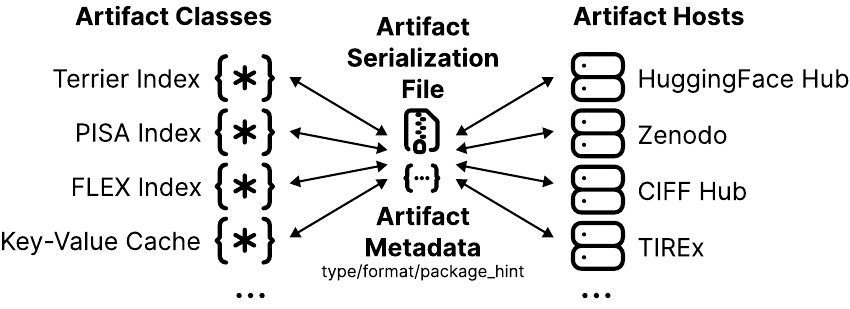}
\vspace{-1em}
\caption{An overview of the distributed design for sharing IR artifacts. Artifacts (instances of Artifact Classes) can be serialized and shared to an Artifact Host via a Artifact Serialization File. Hosted artifacts can then be downloaded and loaded as the original Artifact Class.}
\label{fig:artifact-overview}
\end{figure}

There have been several attempts to improve the sharing and reuse of artifacts in IR. For example, PyTerrier~\cite{DBLP:conf/cikm/MacdonaldTMO21} and Pyserini~\cite{DBLP:conf/sigir/LinMLYPN21} platforms include repositories for pre-built indexes\footnote{\url{http://data.terrier.org/} and \url{https://github.com/castorini/pyserini/blob/master/docs/prebuilt-indexes.md}}. However, these repositories are not publicly writeable, limiting their contents to that of the platform maintainers rather than the IR community at large. TIREx~\cite{DBLP:conf/sigir/FrobeRMDRB0HP23} provides access to built indexes and is built around taking community contributions, but requires that the data generation and indexing code be containerized in Docker images. While this has major benefits in terms of reproducibility, it reduces flexibility by adding the overhead of containerization and makes it more appropriate for sharing final products rather than works-in-progress. BM25S~\cite{DBLP:journals/corr/abs-2407-03618} allows its built indexes to be shared on HuggingFace Hub, but is limited by only supporting its index format and the HuggingFace Hub platform for sharing. CIFF~\cite{DBLP:conf/sigir/LinMKMMSTV20} provides a common exchange format across various inverted index formats and a centralized location for sharing them (CIFF Hub\footnote{\url{https://github.com/pisa-engine/ciff-hub}}), but it is limited to inverted indexes and uploads are limited to the platform maintainers.

The system described in this demonstration paper aims to address the limitations of existing approaches for sharing artifacts in IR by allowing all researchers to share (publicly or privately) any artifact with minimal overhead. The system's design is distributed in nature, with support for virtually any class of artifact (Section~\ref{sec:classes}) and hosting provider (Section~\ref{sec:hosts}), registered via Python packages. The exchange of artifacts is facilitated by a flexible artifact serialization file format (Section~\ref{sec:file}) that provides basic metadata about the type/format of the artifact alongside suggested software to process it. An overview of the design is given in Figure~\ref{fig:artifact-overview}. In demonstration of the flexibility of the format, the system currently supports 13 classes of artifacts (with artifact implementations written in Python, Java, C++, and Rust), and 7 hosts (including HuggingFace Hub, Zenodo, and CIFF Hub). This nearly completely covers (and expands upon) the aforementioned existing artifact-sharing mechanisms.

Several factors have coincided to make this the "right time" to improve the ease of sharing artifacts in IR research.
First, the computational demands of conducting IR research are increasing. With that, it's both more convenient and more environmentally responsible to build off computations that others have done when possible~\cite{DBLP:conf/sigir/ScellsZZ22}.
Second, there is an increasing interest within the community on reproduciblity~\cite{DBLP:conf/sigir/Clancy0HLSW19}. Sharing artifacts enables other researchers to validate intermediate results and assess the reproducibility of different individual components of a research procedure~\cite{DBLP:conf/sigir/WangMMO22}.
Finally, the field is arguably growing more collaborative~\cite{DBLP:journals/sigir/Voorhees20,DBLP:journals/sigir/MacAvaneyRLPEKMFYSFSAFB24}, which increases the need to share artifacts across institutions in the course of day-to-day research.

The system is integrated into PyTerrier~\cite{DBLP:conf/cikm/MacdonaldTMO21}. PyTerrier is a natural platform for the system given its extension-oriented design that provides interoperable functionality across numerous established and emerging IR techniques (e.g.,~\cite{DBLP:conf/sigir/MacAvaneyM22,DBLP:journals/corr/abs-2412-05339,DBLP:conf/ecir/MerkerBFHSWSHP25,conf/sigir/Macdonald25rag}). However, the design of the system described in this paper is not dependent on PyTerrier, and we welcome contributions that apply the design in other toolkits since doing so would help further broaden artifact sharing and reuse in the field. We have already used this system to share artifacts for a published research paper~\cite{DBLP:conf/sigir/ChangMMM24} and to help facilitate cross-institutional collaborations~\cite{DBLP:conf/doceng/KulkarniGFM24,DBLP:conf/wsdm/RatheeMA25,DBLP:conf/ecir/RatheeMA25,conf/sigir/Rathee25telescope}. By providing this demonstration, we hope to help encourage more artifact sharing and ease more collaborations.

\section{System at a Glance}

The artifact-sharing system is geared towards IR researchers, especially those looking to conduct collaborative research (e.g., participate with one another for a shared task) and those looking to share the research artifacts upon publication. Core functionality of the system is demonstrated in a Google Colab notebook,\footnote{Link on the first page of this paper.} which will be made available for SIGIR attendees to interact with on a laptop during the demonstration period. The core functionality is also recorded in the remainder of this section.

The main functionality of the system bundled with the core PyTerrier Python package, which can be installed as follows:

\begin{verbatim}
pip install python-terrier
\end{verbatim}

Artifacts can be loaded from a variety of hosts. For instance, the following code downloads and loads a Terrier~\cite{DBLP:conf/ecir/OunisAPHMJ05} index from the HuggingFace Hub.\footnote{All available artifacts on the HuggingFace Hub can be found through a tag search: \url{https://huggingface.co/datasets?other=pyterrier-artifact}} Once it's loaded, it is ready to use for retrieval:

\begin{small}
\begin{verbatim}
import pyterrier as pt
# load a TerrierIndex object for the msmarco-passage corpus
index = pt.Artifact.from_hf('macavaney/msmarco-passage.terrier')
# the artifact is ready to use:
retriever = index.bm25()
retriever.search('my dear watson')
# qid           query    docno      score  rank
#   1  my dear watson  5341214  36.087756     0
#   1  my dear watson  2385137  30.109050     1
# ...
\end{verbatim}
\end{small}

Note that the system automatically detects the type of artifact and returns an instance of an artifact class that can process it. For instance, if we requested a PISA index~\cite{DBLP:conf/sigir/MalliaSMS19}, it will be loaded as that class (from the pyterrier-pisa~\cite{DBLP:conf/sigir/MacAvaneyM22} extension package\footnote{If the package isn't installed, an error will be raised that gives a hint about what package needs to be installed.}) instead:

\begin{small}
\begin{verbatim}
# load a PisaIndex object
index = pt.Artifact.from_hf('macavaney/msmarco-passage.pisa')
retriever = index.bm25()
retriever.search('my dear watson')
# qid           query    docno      score  rank
#   1  my dear watson  5341214  21.715158     0
#   1  my dear watson  2385137  18.506784     1
# ...
\end{verbatim}
\end{small}
(The results PISA returns for the query are different than Terrier due to differences in tokenization and BM25 score calculations.)

Next, we consider the case where a researcher is looking to share a built artifact (e.g., they are participating in a shared task and built a resource that might be useful for others.) The demonstration shows how you can load a dataset (in this case, a corpus of text from Sir Arthur Conan Doyle, hosted on HuggingFace datasets), index it, and share the artifact to the HuggingFace Hub:

\begin{small}
\begin{verbatim}
from datasets import load_dataset
my_corpus = load_dataset('macavaney/arthur-conan-doyle')['corpus']
index = pt.terrier.TerrierIndex("my-index.terrier")
index.index(corpus)
index.to_hf('my-username/my-index.terrier')
\end{verbatim}
\end{small}

After the upload is complete, the artifact is available for others to using \texttt{Artifact.from\_hf('my-username/my-index.terrier')}.

This demonstration only scratches the surface of the system's available functionality, which is covered in further detail in the following section. Nonetheless, it shows how the core functionality can be useful for those looking to build off somebody else's artifacts and for those looking to easily share artifacts with others.

\section{Technical Details}

There are three core elements to the design of this artifact-sharing system. At its core, the Artifact Serialization File and metadata (Section~\ref{sec:file}) allow artifacts to be represented as a file that includes basic information about the type of artifact stored and suggested software for how to load it. These files can be downloaded from (or, in some cases, uploaded to) Artifact Hosts (Section~\ref{sec:hosts}), which are responsible for the long-term file hosting of the artifact, plus auxiliary features (e.g., search/discoverability, access control, etc.) Finally, Artifact Classes (Section~\ref{sec:classes}) define a Python interface for using the artifacts once downloaded.

Several principles guided the design of this system:

\begin{itemize}
\item \textbf{Flexibility.} The system should be able to support constraints from virtually any artifact class or host---including constraints like the maximum size of individual files (for hosts) or the implementation programming language (for classes).
\item \textbf{Extendable and Decentralized.} The system should be able to be extended with new artifact classes and hosts without needing to change the core package. This is accomplished by defining extensions through Python packages that register their functionality through Python Entry Points.\footnote{\url{https://packaging.python.org/en/latest/specifications/entry-points/}}
\item \textbf{Usability.} Downloaded artifacts should be immediately ready for use through a convenient Python interface.
\end{itemize}

\subsection{Artifact Serialization File and Metadata}\label{sec:file}

The Artifact Serialization File format enables any artifact to be represented as a single file for distribution. At its core, the format is a simple compressed TAR file containing the files that represent the artifact. Many artifacts are already in a binary format that does not benefit much from general compression algorithms, so LZ4 compression was chosen for its reasonably strong compression ratio and its fast compression/decompression rates. The TAR file may include attributes, such as file creation timestamps, though these attributes are removed by default. Similarly, the artifact files can be compressed in any order but are sorted lexically by default to help ensure that functionally-identical artifacts are encoded to the same serialization file. 

\vspace{0.2em}\noindent\textbf{Metadata.} The serialization file should (but does not need to) include a file (\texttt{pt\_meta.json}) that contains JSON-encoded metadata about the artifact. If present, two fields are required: ``type'' (which refers to the broad type of artifact, such as \texttt{sparse\_index}), and ``format'' (which refers to the specific format of the data, such as \texttt{terrier}). Together, this metadata allows the system to identify Artifact Classes that support the type and format of the data. An optional field ``package\_hint'' provides a Python package name to show the user if no Artifact Class is found that supports the type/format. This can aid the user in installing any additional software that they need to use the artifact. Any other metadata may be stored in this file for use by the artifact; other fields are ignored by this system but made available to the loaded artifacts.

Sometimes, the software that creates an artifact may not include the metadata file. For instance, external software may create the artifact, or the artifact may have been created before the introduction of this system. In these case, \texttt{metadata-adapter} entry points allow the system to match an artifact to its corresponding metadata by inspecting its contents. For instance, the \texttt{anserini} adapter checks for the presence of files that suggest that the artifact is an Anserini index. Or, the \texttt{ciff} adapter checks whether the file has a \texttt{.ciff} file extension. Naturally, these checks are potentially error-prone and are more expensive than loading the metadata from a file, so these are only applied as a fallback when the metadata is unavailable.

\vspace{0.2em}\noindent\textbf{Segmented Serialization Files.} Some Artifact Hosts impose a maximum individual file size. For instance, HuggingFace Hub limits individual file sizes to 50GB\footnote{\url{https://huggingface.co/docs/hub/en/storage-limits}}. In these cases, the artifact serialization file can be split into segments with numeric suffixes (\texttt{artifact.tar.lz4.0}, \texttt{artifact.tar.lz4.1}, ...). When the file is segmented, a \texttt{artifact.tar.lz4.json} file\footnote{\href{https://huggingface.co/datasets/macavaney/msmarco-passage.splade-lg.cache/resolve/main/artifact.tar.lz4.json}{example artifact.tar.lz4.json file}} \texttt{must} accompany the files, which provides information about how many segments to expect (among other file metadata, like a checksum). When the artifact is not segmented, the file may be present, but is not required.

\subsection{Artifact Classes}\label{sec:classes}

\begin{table}
\centering
\caption{Examples of Artifact Classes.}
\vspace{-1em}
\scalebox{0.9}{
\begin{tabular}{lll}
\toprule
Class & Type/Format & Package \\
\midrule
\bf Sparse Indexes \\
TerrierIndex & sparse\_index/terrier & \href{https://github.com/terrier-org/pyterrier}{python-terrier} \\
AnseriniIndex & sparse\_index/anserini & \href{https://github.com/seanmacavaney/pyterrier-anserini}{pyterrier-anserini} \\
PisaIndex & sparse\_index/pisa & \href{https://github.com/terrierteam/pyterrier_pisa}{pyterrier-pisa} \\
CiffIndex & sparse\_index/ciff & \href{https://github.com/seanmacavaney/pyterrier-ciff}{pyterrier-ciff} \\
BmpIndex & sparse\_index/bmp & \href{https://github.com/pisa-engine/bmp}{bmp} \\
\midrule
\bf Dense Indexes \\
FlexIndex & dense\_index/flex & \href{https://github.com/terrierteam/pyterrier_dr}{pyterrier-dr} \\
\midrule
\multicolumn{3}{l}{\bf Auxiliary Index Structures} \\
CorpusGraph & corpus\_graph/np\_topk & \href{https://github.com/terrierteam/pyterrier_adaptive}{pyterrier-adaptive} \\
\midrule
\bf Caches \\
KeyValueCache & key\_value\_cache/sqlite3 & \href{https://github.com/terrierteam/pyterrier-caching}{pyterrier-caching} \\
IndexerCache & indexer\_cache/lz4pickle & \href{https://github.com/terrierteam/pyterrier-caching}{pyterrier-caching} \\
RetrieverCache & retriever\_cache/dbm.dumb & \href{https://github.com/terrierteam/pyterrier-caching}{pyterrier-caching} \\
ScorerCache & scorer\_cache/sqlite3 & \href{https://github.com/terrierteam/pyterrier-caching}{pyterrier-caching} \\
DenseScorerCache & scorer\_cache/hdf5 & \href{https://github.com/terrierteam/pyterrier-caching}{pyterrier-caching} \\
CDECache & cde\_cache/np\_pickle & \href{https://github.com/terrierteam/pyterrier_dr}{pyterrier-dr} \\
QualCache & quality\_score\_cache/numpy & \href{https://github.com/terrierteam/pyterrier-quality}{pyterrier-quality} \\
\bottomrule
\end{tabular}
}
\label{tab:classes}
\end{table}

To meet the usability design goal, Artifact Classes define a high-level interface to the functionality provided by the artifact. An Artifact Classes is a simple Python class:

\begin{small}
\begin{verbatim}
class MyArtifact(pt.Artifact):
    def retriever(self):
        ... # return an object that retrieves over this index
\end{verbatim}
\end{small}
To ``register'' the class, it should be included as \texttt{artifact} entry point, with a key containing the type and format. This design allows the system to import only the necessary packages.

\begin{table*}
\centering
\caption{Examples of uploading and downloading from various Artifact Hosts.}
\vspace{-1em}
\scalebox{0.95}{
\begin{tabular}{lll}
\toprule
Platform & Upload Example & Download Example \\
\midrule
\href{https://huggingface.co/}{HuggingFace Hub}
    & \texttt{artifact.to\_hf(\textquotesingle user/repo\textquotesingle)}
    & \texttt{Artifact.from\_hf(\textquotesingle user/repo\textquotesingle)}
\\
\href{https://zenodo.org/}{Zenodo}
    & \texttt{artifact.to\_zenodo()}
    & \texttt{Artifact.from\_zenodo(\textquotesingle zenodo-id\textquotesingle)}
\\
\href{https://magic-wormhole.readthedocs.io/}{Magic-Wormhole}
    & \texttt{artifact.to\_p2p()}
    & \texttt{Artifact.from\_p2p(\textquotesingle code\textquotesingle, \textquotesingle /path/to/index\textquotesingle)}
\\
\href{http://data.terrier.org/}{PyTerrier Data Repository}
    & \textit{N/A}
    & \texttt{Artifact.from\_dataset(\textquotesingle dataset-id\textquotesingle, \textquotesingle index-id\textquotesingle)}
\\
\href{https://github.com/pisa-engine/ciff-hub}{CIFF Hub}
    & \textit{N/A}
    & \texttt{Artifact.from\_url(\textquotesingle ciff-hub:ciff-id\textquotesingle)} \textit{(via \href{https://github.com/seanmacavaney/pyterrier-ciff}{pyterrier-ciff})}
\\
\href{https://github.com/castorini/pyserini/blob/master/docs/prebuilt-indexes.md}{Pyserini Prebuilt Indexes}
    & \textit{N/A}
    & \texttt{Artifact.from\_url(\textquotesingle anserini:index-name\textquotesingle)} \textit{(via \href{https://github.com/seanmacavaney/pyterrier-anserini}{pyterrier-anserini})}
\\
\href{https://www.tira.io/task-overview/ir-benchmarks}{TIREx}
    & \textit{N/A}
    & \texttt{Artifact.from\_url(\textquotesingle tira:dataset-id/team/approach \textquotesingle)} \textit{(via \href{https://github.com/tira-io/tira}{tira})}
\\
\midrule
\textit{Any URL}
    & \textit{N/A}
    & \texttt{Artifact.from\_url(\textquotesingle https://domain.com/artifact.tar.lz4\textquotesingle)}
\\
\bottomrule
\end{tabular}
}
\label{tab:platforms}
\end{table*}

The Artifact Class may also include fields named \texttt{ARTIFACT\_TYPE} and \texttt{ARTIFACT\_FORMAT}, which the system can use to help automatically generate the \texttt{pt\_meta.json} file. The \texttt{pt.Artifact} base class provides a variety of functionality, including methods for downloading artifacts, uploading artifacts, and building the serialization file. Additional utilities for working with Artifact Classes are available in the pyterrier-alpha package\footnote{\url{https://github.com/seanmacavaney/pyterrier-alpha}}, including tools to aid in the construction of artifacts on disk and inspection of artifact objects. These utilities may be added to the core package in the future.

The system already supports several artifact classes, which are summarized in Table~\ref{tab:classes}. There are five classes of sparse indexes, including general-purpose Terrier~\cite{DBLP:conf/ecir/OunisAPHMJ05}, Anserini~\cite{DBLP:journals/jdiq/YangFL18}, and PISA~\cite{DBLP:conf/sigir/MalliaSMS19} classes, plus one for CIFF~\cite{DBLP:conf/sigir/LinMKMMSTV20} (a sparse index interchange format) and BMP~\cite{DBLP:conf/sigir/MalliaST24} (a specialized engine for Learned Sparse Retrieval~\cite{DBLP:conf/ecir/NguyenMY23}). The FLEX (FLexible EXecution) Index supports and unifies a variety of dense retrieval engines, including FAISS~\cite{DBLP:journals/corr/abs-2401-08281}, SCANN~\cite{DBLP:conf/icml/GuoSLGSCK20}, LADR~\cite{DBLP:conf/sigir/KulkarniMGF23}, and FlatNav~\cite{DBLP:journals/corr/abs-2412-01940}. We also plan to add dedicated Artifact Classes for these backends in the future. Finally, other classes are available for supporting corpus graph traversal~\cite{DBLP:conf/cikm/MacAvaneyTM22}, specialized caches for contextual document embeddings~\cite{DBLP:journals/corr/abs-2410-02525} and document quality scores~\cite{DBLP:conf/sigir/ChangMMM24}, and general-purpose caches~\cite{DBLP:conf/ecir/MacAvaney25cache}. Naturally, the functionality provided by each Artifact Class will vary depending on the purpose of the artifact; documentation for each artifact is available in the PyTerrier documentation.\footnote{\url{https://pyterrier.readthedocs.io/}}

In summary, Artifact Classes allow users to define the high-level functionality that an artifact provides, allowing users to immediately make use of artifacts after they are downloaded. A variety of Artifact Classes are already defined (with backends written in several languages, including Python, Java, C++, and Rust), and more can be added through extension packages.

\subsection{Artifact Hosts}\label{sec:hosts}

To avoid relying on a single platform for the distribution of artifacts, the system integrates with several existing hosts. The system can also be extended to support new hosts through entry points. This section covers the current hosts available in the core PyTerrier package and extensions. An overview is given in Table~\ref{tab:platforms}.

\vspace{0.2em}\noindent\textbf{HuggingFace Hub.} The HuggingFace Hub\footnote{\url{https://huggingface.co/}} provides researchers with a platform, software, and free storage to share various artifacts. It is most well-known as a place to share trained machine learning models, though it supports other artifacts as well. There are a variety of benefits for choosing to share artifact on the HuggingFace hub. They are highly discoverable via its search and filtering features\footnote{E.g., all artifacts can be found by filtering on the \texttt{pyterrier-artifact} tag: \url{https://huggingface.co/datasets?other=pyterrier-artifact}.}. It also offers high uptime\footnote{\url{https://status.huggingface.co/}} and generous free storage limits\footnote{\url{https://huggingface.co/docs/hub/en/storage-limits}}.

All Artifact Classes have the methods \texttt{.to\_hf(\textquotesingle user/repo\textquotesingle)} and \texttt{.from\_hf(\textquotesingle user/repo\textquotesingle)}, allowing users to both upload and download their built artifacts. The upload process automatically splits the archive into files below the maximum individual file size (using the approach described in Section~\ref{sec:file}), allowing for very large artifacts to be uploaded. Uploading an artifact generates a README file\footnote{Example of README file for a Terrier index for CORD-19~\cite{DBLP:journals/corr/abs-2004-10706} (the corpus for TREC-COVID~\cite{DBLP:journals/sigir/VoorheesABDHLRS20}): \url{https://huggingface.co/datasets/pyterrier/trec-covid.terrier}} that includes placeholders for information that will likely be useful to other researchers, such as an example usage script, benchmarks, and code to reproduce the artifact.

In line with other integrations, the HuggingFace Hub integration exposes a \texttt{hf:user/repo} ``schema'' that allows artifacts to be downloaded with the \texttt{.from\_url} method. The integration also supports private and gated repositories for those who want to limit the access of their artifacts (e.g., for ongoing work-in-progress or artifacts only shareable under data use agreements).

\vspace{0.2em}\noindent\textbf{Zenodo.} Zenodo~\cite{https://doi.org/10.25495/7gxk-rd71} is a free and dedicated platform for sharing research artifacts with a long-lasting service level agreement.\footnote{\url{https://help.zenodo.org/guides/nih/element4/}} The functionality works similarly to that of HuggingFace Hub, with \texttt{.to\_zenodo()} and \texttt{.from\_zenodo()} methods.

\vspace{0.1em}\noindent\textbf{Magic-Wormhole (Peer-to-Peer).} In some cases, one-off peer-to-peer transfers are preferable to approaches that upload and host artifacts. For instance, two collaborators may want to share an artifact that is a work-in-progress. Magic-Wormhole\footnote{https://magic-wormhole.readthedocs.io/} provides an a way to transfer files between machines, and is built into the core PyTrrier package for performing peer-to-peer artifact transfers.

Due to the nature of peer-to-peer transfers, this integration works slightly differently than the others. The user looking to share an artifact first calls \texttt{artifact.to\_p2p()}, which provides a one-time code for sharing the artifact (e.g., \texttt{66-antenna-transit}). The sharer gives this code to the one receiving it, who then calls \texttt{Artifact.from\_p2p(code, path)}, specifying both the code and the desired path to store the artifact at.

\vspace{0.2em}\noindent\textbf{Other Hosts} The system also supports other artifact repositories, including the PyTerrier Data Repository~\cite{DBLP:conf/cikm/MacdonaldTMO21}, TIREX~\cite{DBLP:conf/sigir/FrobeRMDRB0HP23}, CIFF Hub~\cite{DBLP:conf/sigir/LinMKMMSTV20}, and Pyserini Prebuilt Indexes~\cite{DBLP:conf/sigir/LinMLYPN21}. These generally work through the registration of custom URL schemes (via entry points), e.g., \texttt{ciff-hub:ciff-id}. PyTerrier uses a \texttt{.from\_dataset()} method to enable backward compatibility with its past functionality. The custom URL schemes can either translate the provided ID to a full HTTP(S) URL, or download the data through another means and return the path of the artifact on the local filesystem. Finally, the system supports providing a serialized artifact at any arbitrary URL. Collectively, these demonstrate a high degree of flexiblity over different possible hosting methods.

\section{Conclusion}

Artifact reuse is an important part of IR research. It can help facilitate collaborations, enable reproduction studies, and reduce the envirionmental impact of doing IR research. This demonstration presented a new system for sharing artifacts, especially pre-built indexes, to support these research activities. Unlike prior efforts in this direction, this system is designed with flexibility at its core, allowing users to share many types of artifacts over many different hosts. This is facilitated by a simple artifact serialization file format and metadata scheme. Once downloaded, artifacts are ready for use in experiments.

The system currently provides over 100 artifacts on HuggingFace datasets, provides 14 Artifact Classes, and has been used to help facilitate a cross-institutional collaboration. We hope that this demonstration at SIGIR will help others share and re-use artifacts throughout the course of their research.

\begin{acks}
I thank Craig Macdonald and Maik Fröbe for their helpful comments and suggestions on this system. I also thank Maik Fröbe and Patrick Stahl for contributing the TIREx integration.
\end{acks}

\bibliographystyle{ACM-Reference-Format}
\bibliography{sample-base}


\begin{thebibliography}{34}


\ifx \showCODEN    \undefined \def \showCODEN     #1{\unskip}     \fi
\ifx \showDOI      \undefined \def \showDOI       #1{#1}\fi
\ifx \showISBNx    \undefined \def \showISBNx     #1{\unskip}     \fi
\ifx \showISBNxiii \undefined \def \showISBNxiii  #1{\unskip}     \fi
\ifx \showISSN     \undefined \def \showISSN      #1{\unskip}     \fi
\ifx \showLCCN     \undefined \def \showLCCN      #1{\unskip}     \fi
\ifx \shownote     \undefined \def \shownote      #1{#1}          \fi
\ifx \showarticletitle \undefined \def \showarticletitle #1{#1}   \fi
\ifx \showURL      \undefined \def \showURL       {\relax}        \fi
\providecommand\bibfield[2]{#2}
\providecommand\bibinfo[2]{#2}
\providecommand\natexlab[1]{#1}
\providecommand\showeprint[2][]{arXiv:#2}

\bibitem[Chang et~al\mbox{.}(2024)]%
        {DBLP:conf/sigir/ChangMMM24}
\bibfield{author}{\bibinfo{person}{Xuejun Chang}, \bibinfo{person}{Debabrata Mishra}, \bibinfo{person}{Craig Macdonald}, {and} \bibinfo{person}{Sean MacAvaney}.} \bibinfo{year}{2024}\natexlab{}.
\newblock \showarticletitle{Neural Passage Quality Estimation for Static Pruning}. In \bibinfo{booktitle}{\emph{Proceedings of the 47th International {ACM} {SIGIR} Conference on Research and Development in Information Retrieval, {SIGIR} 2024, Washington DC, USA, July 14-18, 2024}}, \bibfield{editor}{\bibinfo{person}{Grace~Hui Yang}, \bibinfo{person}{Hongning Wang}, \bibinfo{person}{Sam Han}, \bibinfo{person}{Claudia Hauff}, \bibinfo{person}{Guido Zuccon}, {and} \bibinfo{person}{Yi~Zhang}} (Eds.). \bibinfo{publisher}{{ACM}}, \bibinfo{pages}{174--185}.
\newblock
\urldef\tempurl%
\url{https://doi.org/10.1145/3626772.3657765}
\showDOI{\tempurl}


\bibitem[Clancy et~al\mbox{.}(2019)]%
        {DBLP:conf/sigir/Clancy0HLSW19}
\bibfield{author}{\bibinfo{person}{Ryan Clancy}, \bibinfo{person}{Nicola Ferro}, \bibinfo{person}{Claudia Hauff}, \bibinfo{person}{Jimmy Lin}, \bibinfo{person}{Tetsuya Sakai}, {and} \bibinfo{person}{Ze~Zhong Wu}.} \bibinfo{year}{2019}\natexlab{}.
\newblock \showarticletitle{The {SIGIR} 2019 Open-Source {IR} Replicability Challenge {(OSIRRC} 2019)}. In \bibinfo{booktitle}{\emph{Proceedings of the 42nd International {ACM} {SIGIR} Conference on Research and Development in Information Retrieval, {SIGIR} 2019, Paris, France, July 21-25, 2019}}, \bibfield{editor}{\bibinfo{person}{Benjamin Piwowarski}, \bibinfo{person}{Max Chevalier}, \bibinfo{person}{{\'{E}}ric Gaussier}, \bibinfo{person}{Yoelle Maarek}, \bibinfo{person}{Jian{-}Yun Nie}, {and} \bibinfo{person}{Falk Scholer}} (Eds.). \bibinfo{publisher}{{ACM}}, \bibinfo{pages}{1432--1434}.
\newblock
\urldef\tempurl%
\url{https://doi.org/10.1145/3331184.3331647}
\showDOI{\tempurl}


\bibitem[Dhole(2024)]%
        {DBLP:journals/corr/abs-2412-05339}
\bibfield{author}{\bibinfo{person}{Kaustubh~D. Dhole}.} \bibinfo{year}{2024}\natexlab{}.
\newblock \showarticletitle{PyTerrier-GenRank: The PyTerrier Plugin for Reranking with Large Language Models}.
\newblock \bibinfo{journal}{\emph{CoRR}}  \bibinfo{volume}{abs/2412.05339} (\bibinfo{year}{2024}).
\newblock
\urldef\tempurl%
\url{https://doi.org/10.48550/ARXIV.2412.05339}
\showDOI{\tempurl}
\showeprint[arXiv]{2412.05339}


\bibitem[Douze et~al\mbox{.}(2024)]%
        {DBLP:journals/corr/abs-2401-08281}
\bibfield{author}{\bibinfo{person}{Matthijs Douze}, \bibinfo{person}{Alexandr Guzhva}, \bibinfo{person}{Chengqi Deng}, \bibinfo{person}{Jeff Johnson}, \bibinfo{person}{Gergely Szilvasy}, \bibinfo{person}{Pierre{-}Emmanuel Mazar{\'{e}}}, \bibinfo{person}{Maria Lomeli}, \bibinfo{person}{Lucas Hosseini}, {and} \bibinfo{person}{Herv{\'{e}} J{\'{e}}gou}.} \bibinfo{year}{2024}\natexlab{}.
\newblock \showarticletitle{The Faiss library}.
\newblock \bibinfo{journal}{\emph{CoRR}}  \bibinfo{volume}{abs/2401.08281} (\bibinfo{year}{2024}).
\newblock
\urldef\tempurl%
\url{https://doi.org/10.48550/ARXIV.2401.08281}
\showDOI{\tempurl}
\showeprint[arXiv]{2401.08281}


\bibitem[{European Organization For Nuclear Research} and {OpenAIRE}(2013)]%
        {https://doi.org/10.25495/7gxk-rd71}
\bibfield{author}{\bibinfo{person}{{European Organization For Nuclear Research}} {and} \bibinfo{person}{{OpenAIRE}}.} \bibinfo{year}{2013}\natexlab{}.
\newblock \bibinfo{title}{Zenodo}.
\newblock
\newblock
\urldef\tempurl%
\url{https://doi.org/10.25495/7GXK-RD71}
\showDOI{\tempurl}


\bibitem[Fr{\"{o}}be et~al\mbox{.}(2023)]%
        {DBLP:conf/sigir/FrobeRMDRB0HP23}
\bibfield{author}{\bibinfo{person}{Maik Fr{\"{o}}be}, \bibinfo{person}{Jan~Heinrich Reimer}, \bibinfo{person}{Sean MacAvaney}, \bibinfo{person}{Niklas Deckers}, \bibinfo{person}{Simon Reich}, \bibinfo{person}{Janek Bevendorff}, \bibinfo{person}{Benno Stein}, \bibinfo{person}{Matthias Hagen}, {and} \bibinfo{person}{Martin Potthast}.} \bibinfo{year}{2023}\natexlab{}.
\newblock \showarticletitle{The Information Retrieval Experiment Platform}. In \bibinfo{booktitle}{\emph{Proceedings of the 46th International {ACM} {SIGIR} Conference on Research and Development in Information Retrieval, {SIGIR} 2023, Taipei, Taiwan, July 23-27, 2023}}, \bibfield{editor}{\bibinfo{person}{Hsin{-}Hsi Chen}, \bibinfo{person}{Wei{-}Jou~(Edward) Duh}, \bibinfo{person}{Hen{-}Hsen Huang}, \bibinfo{person}{Makoto~P. Kato}, \bibinfo{person}{Josiane Mothe}, {and} \bibinfo{person}{Barbara Poblete}} (Eds.). \bibinfo{publisher}{{ACM}}, \bibinfo{pages}{2826--2836}.
\newblock
\urldef\tempurl%
\url{https://doi.org/10.1145/3539618.3591888}
\showDOI{\tempurl}


\bibitem[Guo et~al\mbox{.}(2020)]%
        {DBLP:conf/icml/GuoSLGSCK20}
\bibfield{author}{\bibinfo{person}{Ruiqi Guo}, \bibinfo{person}{Philip Sun}, \bibinfo{person}{Erik Lindgren}, \bibinfo{person}{Quan Geng}, \bibinfo{person}{David Simcha}, \bibinfo{person}{Felix Chern}, {and} \bibinfo{person}{Sanjiv Kumar}.} \bibinfo{year}{2020}\natexlab{}.
\newblock \showarticletitle{Accelerating Large-Scale Inference with Anisotropic Vector Quantization}. In \bibinfo{booktitle}{\emph{Proceedings of the 37th International Conference on Machine Learning, {ICML} 2020, 13-18 July 2020, Virtual Event}} \emph{(\bibinfo{series}{Proceedings of Machine Learning Research}, Vol.~\bibinfo{volume}{119})}. \bibinfo{publisher}{{PMLR}}, \bibinfo{pages}{3887--3896}.
\newblock
\urldef\tempurl%
\url{http://proceedings.mlr.press/v119/guo20h.html}
\showURL{%
\tempurl}


\bibitem[Kulkarni et~al\mbox{.}(2024)]%
        {DBLP:conf/doceng/KulkarniGFM24}
\bibfield{author}{\bibinfo{person}{Hrishikesh Kulkarni}, \bibinfo{person}{Nazli Goharian}, \bibinfo{person}{Ophir Frieder}, {and} \bibinfo{person}{Sean MacAvaney}.} \bibinfo{year}{2024}\natexlab{}.
\newblock \showarticletitle{LexBoost: Improving Lexical Document Retrieval with Nearest Neighbors}. In \bibinfo{booktitle}{\emph{Proceedings of the {ACM} Symposium on Document Engineering 2024, DocEng 2024, San Jose, CA, USA, August 20-23, 2024}}. \bibinfo{publisher}{{ACM}}, \bibinfo{pages}{16:1--16:10}.
\newblock
\urldef\tempurl%
\url{https://doi.org/10.1145/3685650.3685658}
\showDOI{\tempurl}


\bibitem[Kulkarni et~al\mbox{.}(2023)]%
        {DBLP:conf/sigir/KulkarniMGF23}
\bibfield{author}{\bibinfo{person}{Hrishikesh Kulkarni}, \bibinfo{person}{Sean MacAvaney}, \bibinfo{person}{Nazli Goharian}, {and} \bibinfo{person}{Ophir Frieder}.} \bibinfo{year}{2023}\natexlab{}.
\newblock \showarticletitle{Lexically-Accelerated Dense Retrieval}. In \bibinfo{booktitle}{\emph{Proceedings of the 46th International {ACM} {SIGIR} Conference on Research and Development in Information Retrieval, {SIGIR} 2023, Taipei, Taiwan, July 23-27, 2023}}, \bibfield{editor}{\bibinfo{person}{Hsin{-}Hsi Chen}, \bibinfo{person}{Wei{-}Jou~(Edward) Duh}, \bibinfo{person}{Hen{-}Hsen Huang}, \bibinfo{person}{Makoto~P. Kato}, \bibinfo{person}{Josiane Mothe}, {and} \bibinfo{person}{Barbara Poblete}} (Eds.). \bibinfo{publisher}{{ACM}}, \bibinfo{pages}{152--162}.
\newblock
\urldef\tempurl%
\url{https://doi.org/10.1145/3539618.3591715}
\showDOI{\tempurl}


\bibitem[Lin et~al\mbox{.}(2021)]%
        {DBLP:conf/sigir/LinMLYPN21}
\bibfield{author}{\bibinfo{person}{Jimmy Lin}, \bibinfo{person}{Xueguang Ma}, \bibinfo{person}{Sheng{-}Chieh Lin}, \bibinfo{person}{Jheng{-}Hong Yang}, \bibinfo{person}{Ronak Pradeep}, {and} \bibinfo{person}{Rodrigo~Frassetto Nogueira}.} \bibinfo{year}{2021}\natexlab{}.
\newblock \showarticletitle{Pyserini: {A} Python Toolkit for Reproducible Information Retrieval Research with Sparse and Dense Representations}. In \bibinfo{booktitle}{\emph{{SIGIR} '21: The 44th International {ACM} {SIGIR} Conference on Research and Development in Information Retrieval, Virtual Event, Canada, July 11-15, 2021}}, \bibfield{editor}{\bibinfo{person}{Fernando Diaz}, \bibinfo{person}{Chirag Shah}, \bibinfo{person}{Torsten Suel}, \bibinfo{person}{Pablo Castells}, \bibinfo{person}{Rosie Jones}, {and} \bibinfo{person}{Tetsuya Sakai}} (Eds.). \bibinfo{publisher}{{ACM}}, \bibinfo{pages}{2356--2362}.
\newblock
\urldef\tempurl%
\url{https://doi.org/10.1145/3404835.3463238}
\showDOI{\tempurl}


\bibitem[Lin et~al\mbox{.}(2020)]%
        {DBLP:conf/sigir/LinMKMMSTV20}
\bibfield{author}{\bibinfo{person}{Jimmy Lin}, \bibinfo{person}{Joel~M. Mackenzie}, \bibinfo{person}{Chris Kamphuis}, \bibinfo{person}{Craig Macdonald}, \bibinfo{person}{Antonio Mallia}, \bibinfo{person}{Michal Siedlaczek}, \bibinfo{person}{Andrew Trotman}, {and} \bibinfo{person}{Arjen~P. de Vries}.} \bibinfo{year}{2020}\natexlab{}.
\newblock \showarticletitle{Supporting Interoperability Between Open-Source Search Engines with the Common Index File Format}. In \bibinfo{booktitle}{\emph{Proceedings of the 43rd International {ACM} {SIGIR} conference on research and development in Information Retrieval, {SIGIR} 2020, Virtual Event, China, July 25-30, 2020}}, \bibfield{editor}{\bibinfo{person}{Jimmy~X. Huang}, \bibinfo{person}{Yi~Chang}, \bibinfo{person}{Xueqi Cheng}, \bibinfo{person}{Jaap Kamps}, \bibinfo{person}{Vanessa Murdock}, \bibinfo{person}{Ji{-}Rong Wen}, {and} \bibinfo{person}{Yiqun Liu}} (Eds.). \bibinfo{publisher}{{ACM}}, \bibinfo{pages}{2149--2152}.
\newblock
\urldef\tempurl%
\url{https://doi.org/10.1145/3397271.3401404}
\showDOI{\tempurl}


\bibitem[L{\`{u}}(2024)]%
        {DBLP:journals/corr/abs-2407-03618}
\bibfield{author}{\bibinfo{person}{Xing~Han L{\`{u}}}.} \bibinfo{year}{2024}\natexlab{}.
\newblock \showarticletitle{{BM25S:} Orders of magnitude faster lexical search via eager sparse scoring}.
\newblock \bibinfo{journal}{\emph{CoRR}}  \bibinfo{volume}{abs/2407.03618} (\bibinfo{year}{2024}).
\newblock
\urldef\tempurl%
\url{https://doi.org/10.48550/ARXIV.2407.03618}
\showDOI{\tempurl}
\showeprint[arXiv]{2407.03618}


\bibitem[MacAvaney and Macdonald(2022)]%
        {DBLP:conf/sigir/MacAvaneyM22}
\bibfield{author}{\bibinfo{person}{Sean MacAvaney} {and} \bibinfo{person}{Craig Macdonald}.} \bibinfo{year}{2022}\natexlab{}.
\newblock \showarticletitle{A Python Interface to PISA!}. In \bibinfo{booktitle}{\emph{{SIGIR} '22: The 45th International {ACM} {SIGIR} Conference on Research and Development in Information Retrieval, Madrid, Spain, July 11 - 15, 2022}}, \bibfield{editor}{\bibinfo{person}{Enrique Amig{\'{o}}}, \bibinfo{person}{Pablo Castells}, \bibinfo{person}{Julio Gonzalo}, \bibinfo{person}{Ben Carterette}, \bibinfo{person}{J.~Shane Culpepper}, {and} \bibinfo{person}{Gabriella Kazai}} (Eds.). \bibinfo{publisher}{{ACM}}, \bibinfo{pages}{3339--3344}.
\newblock
\urldef\tempurl%
\url{https://doi.org/10.1145/3477495.3531656}
\showDOI{\tempurl}


\bibitem[MacAvaney and Macdonald(2025)]%
        {DBLP:conf/ecir/MacAvaney25cache}
\bibfield{author}{\bibinfo{person}{Sean MacAvaney} {and} \bibinfo{person}{Craig Macdonald}.} \bibinfo{year}{2025}\natexlab{}.
\newblock \showarticletitle{On Precomputation and Caching in Information Retrieval Experiments with Pipeline Architectures}. In \bibinfo{booktitle}{\emph{Proceedings of the 2nd International Workshop on Open Web Search ({WOWS}) co-located with 47th European Conference on Information Retrieval, {ECIR} 2025, Lucca, Italy, April 6-10, 2025}} \emph{(\bibinfo{series}{{CEUR} Workshop Proceedings})}. \bibinfo{publisher}{CEUR-WS.org}.
\newblock


\bibitem[MacAvaney et~al\mbox{.}(2024)]%
        {DBLP:journals/sigir/MacAvaneyRLPEKMFYSFSAFB24}
\bibfield{author}{\bibinfo{person}{Sean MacAvaney}, \bibinfo{person}{Adam Roegiest}, \bibinfo{person}{Aldo Lipani}, \bibinfo{person}{Andrew Parry}, \bibinfo{person}{Bj{\"{o}}rn Engelmann}, \bibinfo{person}{Christin~Katharina Kreutz}, \bibinfo{person}{Chuan Meng}, \bibinfo{person}{Erlend Frayling}, \bibinfo{person}{Eugene Yang}, \bibinfo{person}{Ferdinand Schlatt}, \bibinfo{person}{Guglielmo Faggioli}, \bibinfo{person}{Harrisen Scells}, \bibinfo{person}{Iana Atanassova}, \bibinfo{person}{Jana Friese}, \bibinfo{person}{Janek Bevendorff}, \bibinfo{person}{Javier Sanz{-}Cruzado}, \bibinfo{person}{Johanne Trippas}, \bibinfo{person}{Kanaad Pathak}, \bibinfo{person}{Kaustubh~D. Dhole}, \bibinfo{person}{Leif Azzopardi}, \bibinfo{person}{Maik Fr{\"{o}}be}, \bibinfo{person}{Marc Bertin}, \bibinfo{person}{Nishchal Prasad}, \bibinfo{person}{Saber Zerhoudi}, \bibinfo{person}{Shuai Wang}, \bibinfo{person}{Shubham Chatterjee}, \bibinfo{person}{Thomas J{\"{a}}nich}, \bibinfo{person}{Udo Kruschwitz}, \bibinfo{person}{Xi
  Wang}, {and} \bibinfo{person}{Zijun Long}.} \bibinfo{year}{2024}\natexlab{}.
\newblock \showarticletitle{Report on the Collab-a-Thon at {ECIR} 2024}.
\newblock \bibinfo{journal}{\emph{{SIGIR} Forum}} \bibinfo{volume}{58}, \bibinfo{number}{1} (\bibinfo{year}{2024}), \bibinfo{pages}{1--11}.
\newblock
\urldef\tempurl%
\url{https://doi.org/10.1145/3687273.3687287}
\showDOI{\tempurl}


\bibitem[MacAvaney et~al\mbox{.}(2022)]%
        {DBLP:conf/cikm/MacAvaneyTM22}
\bibfield{author}{\bibinfo{person}{Sean MacAvaney}, \bibinfo{person}{Nicola Tonellotto}, {and} \bibinfo{person}{Craig Macdonald}.} \bibinfo{year}{2022}\natexlab{}.
\newblock \showarticletitle{Adaptive Re-Ranking with a Corpus Graph}. In \bibinfo{booktitle}{\emph{Proceedings of the 31st {ACM} International Conference on Information {\&} Knowledge Management, Atlanta, GA, USA, October 17-21, 2022}}, \bibfield{editor}{\bibinfo{person}{Mohammad~Al Hasan} {and} \bibinfo{person}{Li~Xiong}} (Eds.). \bibinfo{publisher}{{ACM}}, \bibinfo{pages}{1491--1500}.
\newblock
\urldef\tempurl%
\url{https://doi.org/10.1145/3511808.3557231}
\showDOI{\tempurl}


\bibitem[Macdonald et~al\mbox{.}(2025)]%
        {conf/sigir/Macdonald25rag}
\bibfield{author}{\bibinfo{person}{Craig Macdonald}, \bibinfo{person}{Jinyuan Fang}, \bibinfo{person}{Andrew Parry}, \bibinfo{person}{Craig Macdonald}, {and} \bibinfo{person}{Zaiqiao Meng}.} \bibinfo{year}{2025}\natexlab{}.
\newblock \showarticletitle{Constructing and Evaluating Declarative {RAG} Pipelines in {PyTerrier}}. In \bibinfo{booktitle}{\emph{Proceedings of the 48th International {ACM} {SIGIR} conference on research and development in Information Retrieval, {SIGIR} 2025}}. \bibinfo{publisher}{{ACM}}.
\newblock
\urldef\tempurl%
\url{https://doi.org/10.1145/3726302.3730150}
\showDOI{\tempurl}


\bibitem[Macdonald et~al\mbox{.}(2021)]%
        {DBLP:conf/cikm/MacdonaldTMO21}
\bibfield{author}{\bibinfo{person}{Craig Macdonald}, \bibinfo{person}{Nicola Tonellotto}, \bibinfo{person}{Sean MacAvaney}, {and} \bibinfo{person}{Iadh Ounis}.} \bibinfo{year}{2021}\natexlab{}.
\newblock \showarticletitle{PyTerrier: Declarative Experimentation in Python from {BM25} to Dense Retrieval}. In \bibinfo{booktitle}{\emph{{CIKM} '21: The 30th {ACM} International Conference on Information and Knowledge Management, Virtual Event, Queensland, Australia, November 1 - 5, 2021}}, \bibfield{editor}{\bibinfo{person}{Gianluca Demartini}, \bibinfo{person}{Guido Zuccon}, \bibinfo{person}{J.~Shane Culpepper}, \bibinfo{person}{Zi~Huang}, {and} \bibinfo{person}{Hanghang Tong}} (Eds.). \bibinfo{publisher}{{ACM}}, \bibinfo{pages}{4526--4533}.
\newblock
\urldef\tempurl%
\url{https://doi.org/10.1145/3459637.3482013}
\showDOI{\tempurl}


\bibitem[Mallia et~al\mbox{.}(2019)]%
        {DBLP:conf/sigir/MalliaSMS19}
\bibfield{author}{\bibinfo{person}{Antonio Mallia}, \bibinfo{person}{Michal Siedlaczek}, \bibinfo{person}{Joel~M. Mackenzie}, {and} \bibinfo{person}{Torsten Suel}.} \bibinfo{year}{2019}\natexlab{}.
\newblock \showarticletitle{{PISA:} Performant Indexes and Search for Academia}. In \bibinfo{booktitle}{\emph{Proceedings of the Open-Source {IR} Replicability Challenge co-located with 42nd International {ACM} {SIGIR} Conference on Research and Development in Information Retrieval, OSIRRC@SIGIR 2019, Paris, France, July 25, 2019}} \emph{(\bibinfo{series}{{CEUR} Workshop Proceedings}, Vol.~\bibinfo{volume}{2409})}, \bibfield{editor}{\bibinfo{person}{Ryan Clancy}, \bibinfo{person}{Nicola Ferro}, \bibinfo{person}{Claudia Hauff}, \bibinfo{person}{Jimmy Lin}, \bibinfo{person}{Tetsuya Sakai}, {and} \bibinfo{person}{Ze~Zhong Wu}} (Eds.). \bibinfo{publisher}{CEUR-WS.org}, \bibinfo{pages}{50--56}.
\newblock
\urldef\tempurl%
\url{https://ceur-ws.org/Vol-2409/docker08.pdf}
\showURL{%
\tempurl}


\bibitem[Mallia et~al\mbox{.}(2024)]%
        {DBLP:conf/sigir/MalliaST24}
\bibfield{author}{\bibinfo{person}{Antonio Mallia}, \bibinfo{person}{Torsten Suel}, {and} \bibinfo{person}{Nicola Tonellotto}.} \bibinfo{year}{2024}\natexlab{}.
\newblock \showarticletitle{Faster Learned Sparse Retrieval with Block-Max Pruning}. In \bibinfo{booktitle}{\emph{Proceedings of the 47th International {ACM} {SIGIR} Conference on Research and Development in Information Retrieval, {SIGIR} 2024, Washington DC, USA, July 14-18, 2024}}, \bibfield{editor}{\bibinfo{person}{Grace~Hui Yang}, \bibinfo{person}{Hongning Wang}, \bibinfo{person}{Sam Han}, \bibinfo{person}{Claudia Hauff}, \bibinfo{person}{Guido Zuccon}, {and} \bibinfo{person}{Yi~Zhang}} (Eds.). \bibinfo{publisher}{{ACM}}, \bibinfo{pages}{2411--2415}.
\newblock
\urldef\tempurl%
\url{https://doi.org/10.1145/3626772.3657906}
\showDOI{\tempurl}


\bibitem[Merker et~al\mbox{.}(2025)]%
        {DBLP:conf/ecir/MerkerBFHSWSHP25}
\bibfield{author}{\bibinfo{person}{Jan~Heinrich Merker}, \bibinfo{person}{Janek Bevendorff}, \bibinfo{person}{Maik Fr{\"{o}}be}, \bibinfo{person}{Tim Hagen}, \bibinfo{person}{Harrisen Scells}, \bibinfo{person}{Matti Wiegmann}, \bibinfo{person}{Benno Stein}, \bibinfo{person}{Matthias Hagen}, {and} \bibinfo{person}{Martin Potthast}.} \bibinfo{year}{2025}\natexlab{}.
\newblock \showarticletitle{Web-Scale Retrieval Experimentation with chatnoir-pyterrier}. In \bibinfo{booktitle}{\emph{Advances in Information Retrieval - 47th European Conference on Information Retrieval, {ECIR} 2025, Lucca, Italy, April 6-10, 2025, Proceedings, Part {V}}} \emph{(\bibinfo{series}{Lecture Notes in Computer Science}, Vol.~\bibinfo{volume}{15576})}, \bibfield{editor}{\bibinfo{person}{Claudia Hauff}, \bibinfo{person}{Craig Macdonald}, \bibinfo{person}{Dietmar Jannach}, \bibinfo{person}{Gabriella Kazai}, \bibinfo{person}{Franco~Maria Nardini}, \bibinfo{person}{Fabio Pinelli}, \bibinfo{person}{Fabrizio Silvestri}, {and} \bibinfo{person}{Nicola Tonellotto}} (Eds.). \bibinfo{publisher}{Springer}, \bibinfo{pages}{96--104}.
\newblock
\urldef\tempurl%
\url{https://doi.org/10.1007/978-3-031-88720-8\_17}
\showDOI{\tempurl}


\bibitem[Morris and Rush(2024)]%
        {DBLP:journals/corr/abs-2410-02525}
\bibfield{author}{\bibinfo{person}{John~X. Morris} {and} \bibinfo{person}{Alexander~M. Rush}.} \bibinfo{year}{2024}\natexlab{}.
\newblock \showarticletitle{Contextual Document Embeddings}.
\newblock \bibinfo{journal}{\emph{CoRR}}  \bibinfo{volume}{abs/2410.02525} (\bibinfo{year}{2024}).
\newblock
\urldef\tempurl%
\url{https://doi.org/10.48550/ARXIV.2410.02525}
\showDOI{\tempurl}
\showeprint[arXiv]{2410.02525}


\bibitem[Munyampirwa et~al\mbox{.}(2024)]%
        {DBLP:journals/corr/abs-2412-01940}
\bibfield{author}{\bibinfo{person}{Blaise Munyampirwa}, \bibinfo{person}{Vihan Lakshman}, {and} \bibinfo{person}{Benjamin Coleman}.} \bibinfo{year}{2024}\natexlab{}.
\newblock \showarticletitle{Down with the Hierarchy: The 'H' in {HNSW} Stands for "Hubs"}.
\newblock \bibinfo{journal}{\emph{CoRR}}  \bibinfo{volume}{abs/2412.01940} (\bibinfo{year}{2024}).
\newblock
\urldef\tempurl%
\url{https://doi.org/10.48550/ARXIV.2412.01940}
\showDOI{\tempurl}
\showeprint[arXiv]{2412.01940}


\bibitem[Nguyen et~al\mbox{.}(2023)]%
        {DBLP:conf/ecir/NguyenMY23}
\bibfield{author}{\bibinfo{person}{Thong Nguyen}, \bibinfo{person}{Sean MacAvaney}, {and} \bibinfo{person}{Andrew Yates}.} \bibinfo{year}{2023}\natexlab{}.
\newblock \showarticletitle{A Unified Framework for Learned Sparse Retrieval}. In \bibinfo{booktitle}{\emph{Advances in Information Retrieval - 45th European Conference on Information Retrieval, {ECIR} 2023, Dublin, Ireland, April 2-6, 2023, Proceedings, Part {III}}} \emph{(\bibinfo{series}{Lecture Notes in Computer Science}, Vol.~\bibinfo{volume}{13982})}, \bibfield{editor}{\bibinfo{person}{Jaap Kamps}, \bibinfo{person}{Lorraine Goeuriot}, \bibinfo{person}{Fabio Crestani}, \bibinfo{person}{Maria Maistro}, \bibinfo{person}{Hideo Joho}, \bibinfo{person}{Brian Davis}, \bibinfo{person}{Cathal Gurrin}, \bibinfo{person}{Udo Kruschwitz}, {and} \bibinfo{person}{Annalina Caputo}} (Eds.). \bibinfo{publisher}{Springer}, \bibinfo{pages}{101--116}.
\newblock
\urldef\tempurl%
\url{https://doi.org/10.1007/978-3-031-28241-6\_7}
\showDOI{\tempurl}


\bibitem[Ounis et~al\mbox{.}(2005)]%
        {DBLP:conf/ecir/OunisAPHMJ05}
\bibfield{author}{\bibinfo{person}{Iadh Ounis}, \bibinfo{person}{Gianni Amati}, \bibinfo{person}{Vassilis Plachouras}, \bibinfo{person}{Ben He}, \bibinfo{person}{Craig Macdonald}, {and} \bibinfo{person}{Douglas Johnson}.} \bibinfo{year}{2005}\natexlab{}.
\newblock \showarticletitle{Terrier Information Retrieval Platform}. In \bibinfo{booktitle}{\emph{Advances in Information Retrieval, 27th European Conference on {IR} Research, {ECIR} 2005, Santiago de Compostela, Spain, March 21-23, 2005, Proceedings}} \emph{(\bibinfo{series}{Lecture Notes in Computer Science}, Vol.~\bibinfo{volume}{3408})}, \bibfield{editor}{\bibinfo{person}{David~E. Losada} {and} \bibinfo{person}{Juan~M. Fern{\'{a}}ndez{-}Luna}} (Eds.). \bibinfo{publisher}{Springer}, \bibinfo{pages}{517--519}.
\newblock
\urldef\tempurl%
\url{https://doi.org/10.1007/978-3-540-31865-1\_37}
\showDOI{\tempurl}


\bibitem[Rathee et~al\mbox{.}(2025a)]%
        {DBLP:conf/ecir/RatheeMA25}
\bibfield{author}{\bibinfo{person}{Mandeep Rathee}, \bibinfo{person}{Sean MacAvaney}, {and} \bibinfo{person}{Avishek Anand}.} \bibinfo{year}{2025}\natexlab{a}.
\newblock \showarticletitle{Guiding Retrieval Using LLM-Based Listwise Rankers}. In \bibinfo{booktitle}{\emph{Advances in Information Retrieval - 47th European Conference on Information Retrieval, {ECIR} 2025, Lucca, Italy, April 6-10, 2025, Proceedings, Part {I}}} \emph{(\bibinfo{series}{Lecture Notes in Computer Science}, Vol.~\bibinfo{volume}{15572})}, \bibfield{editor}{\bibinfo{person}{Claudia Hauff}, \bibinfo{person}{Craig Macdonald}, \bibinfo{person}{Dietmar Jannach}, \bibinfo{person}{Gabriella Kazai}, \bibinfo{person}{Franco~Maria Nardini}, \bibinfo{person}{Fabio Pinelli}, \bibinfo{person}{Fabrizio Silvestri}, {and} \bibinfo{person}{Nicola Tonellotto}} (Eds.). \bibinfo{publisher}{Springer}, \bibinfo{pages}{230--246}.
\newblock
\urldef\tempurl%
\url{https://doi.org/10.1007/978-3-031-88708-6\_15}
\showDOI{\tempurl}


\bibitem[Rathee et~al\mbox{.}(2025b)]%
        {DBLP:conf/wsdm/RatheeMA25}
\bibfield{author}{\bibinfo{person}{Mandeep Rathee}, \bibinfo{person}{Sean MacAvaney}, {and} \bibinfo{person}{Avishek Anand}.} \bibinfo{year}{2025}\natexlab{b}.
\newblock \showarticletitle{Quam: Adaptive Retrieval through Query Affinity Modelling}. In \bibinfo{booktitle}{\emph{Proceedings of the Eighteenth {ACM} International Conference on Web Search and Data Mining, {WSDM} 2025, Hannover, Germany, March 10-14, 2025}}, \bibfield{editor}{\bibinfo{person}{Wolfgang Nejdl}, \bibinfo{person}{S{\"{o}}ren Auer}, \bibinfo{person}{Meeyoung Cha}, \bibinfo{person}{Marie{-}Francine Moens}, {and} \bibinfo{person}{Marc Najork}} (Eds.). \bibinfo{publisher}{{ACM}}, \bibinfo{pages}{954--962}.
\newblock
\urldef\tempurl%
\url{https://doi.org/10.1145/3701551.3703584}
\showDOI{\tempurl}


\bibitem[Rathee et~al\mbox{.}(2025c)]%
        {conf/sigir/Rathee25telescope}
\bibfield{author}{\bibinfo{person}{Mandeep Rathee}, \bibinfo{person}{V Venktesh}, \bibinfo{person}{Sean MacAvaney}, {and} \bibinfo{person}{Avishek Anand}.} \bibinfo{year}{2025}\natexlab{c}.
\newblock \showarticletitle{Breaking the Lens of the Telescope: Online Relevance Estimation over Large Retrieval Sets}. In \bibinfo{booktitle}{\emph{Proceedings of the 48th International {ACM} {SIGIR} conference on research and development in Information Retrieval, {SIGIR} 2025}}. \bibinfo{publisher}{{ACM}}.
\newblock
\urldef\tempurl%
\url{https://doi.org/10.1145/3726302.3729910}
\showDOI{\tempurl}


\bibitem[Scells et~al\mbox{.}(2022)]%
        {DBLP:conf/sigir/ScellsZZ22}
\bibfield{author}{\bibinfo{person}{Harrisen Scells}, \bibinfo{person}{Shengyao Zhuang}, {and} \bibinfo{person}{Guido Zuccon}.} \bibinfo{year}{2022}\natexlab{}.
\newblock \showarticletitle{Reduce, Reuse, Recycle: Green Information Retrieval Research}. In \bibinfo{booktitle}{\emph{{SIGIR} '22: The 45th International {ACM} {SIGIR} Conference on Research and Development in Information Retrieval, Madrid, Spain, July 11 - 15, 2022}}, \bibfield{editor}{\bibinfo{person}{Enrique Amig{\'{o}}}, \bibinfo{person}{Pablo Castells}, \bibinfo{person}{Julio Gonzalo}, \bibinfo{person}{Ben Carterette}, \bibinfo{person}{J.~Shane Culpepper}, {and} \bibinfo{person}{Gabriella Kazai}} (Eds.). \bibinfo{publisher}{{ACM}}, \bibinfo{pages}{2825--2837}.
\newblock
\urldef\tempurl%
\url{https://doi.org/10.1145/3477495.3531766}
\showDOI{\tempurl}


\bibitem[Voorhees(2020)]%
        {DBLP:journals/sigir/Voorhees20}
\bibfield{author}{\bibinfo{person}{Ellen~M. Voorhees}.} \bibinfo{year}{2020}\natexlab{}.
\newblock \showarticletitle{Coopetition in {IR} research}.
\newblock \bibinfo{journal}{\emph{{SIGIR} Forum}} \bibinfo{volume}{54}, \bibinfo{number}{2} (\bibinfo{year}{2020}), \bibinfo{pages}{1:1--1:3}.
\newblock
\urldef\tempurl%
\url{https://doi.org/10.1145/3483382.3483384}
\showDOI{\tempurl}


\bibitem[Voorhees et~al\mbox{.}(2020)]%
        {DBLP:journals/sigir/VoorheesABDHLRS20}
\bibfield{author}{\bibinfo{person}{Ellen~M. Voorhees}, \bibinfo{person}{Tasmeer Alam}, \bibinfo{person}{Steven Bedrick}, \bibinfo{person}{Dina Demner{-}Fushman}, \bibinfo{person}{William~R. Hersh}, \bibinfo{person}{Kyle Lo}, \bibinfo{person}{Kirk Roberts}, \bibinfo{person}{Ian Soboroff}, {and} \bibinfo{person}{Lucy~Lu Wang}.} \bibinfo{year}{2020}\natexlab{}.
\newblock \showarticletitle{{TREC-COVID:} constructing a pandemic information retrieval test collection}.
\newblock \bibinfo{journal}{\emph{{SIGIR} Forum}} \bibinfo{volume}{54}, \bibinfo{number}{1} (\bibinfo{year}{2020}), \bibinfo{pages}{1:1--1:12}.
\newblock
\urldef\tempurl%
\url{https://doi.org/10.1145/3451964.3451965}
\showDOI{\tempurl}


\bibitem[Wang et~al\mbox{.}(2020)]%
        {DBLP:journals/corr/abs-2004-10706}
\bibfield{author}{\bibinfo{person}{Lucy~Lu Wang}, \bibinfo{person}{Kyle Lo}, \bibinfo{person}{Yoganand Chandrasekhar}, \bibinfo{person}{Russell Reas}, \bibinfo{person}{Jiangjiang Yang}, \bibinfo{person}{Darrin Eide}, \bibinfo{person}{Kathryn Funk}, \bibinfo{person}{Rodney Kinney}, \bibinfo{person}{Ziyang Liu}, \bibinfo{person}{William Merrill}, \bibinfo{person}{Paul Mooney}, \bibinfo{person}{Dewey~A. Murdick}, \bibinfo{person}{Devvret Rishi}, \bibinfo{person}{Jerry Sheehan}, \bibinfo{person}{Zhihong Shen}, \bibinfo{person}{Brandon Stilson}, \bibinfo{person}{Alex~D. Wade}, \bibinfo{person}{Kuansan Wang}, \bibinfo{person}{Chris Wilhelm}, \bibinfo{person}{Boya Xie}, \bibinfo{person}{Douglas Raymond}, \bibinfo{person}{Daniel~S. Weld}, \bibinfo{person}{Oren Etzioni}, {and} \bibinfo{person}{Sebastian Kohlmeier}.} \bibinfo{year}{2020}\natexlab{}.
\newblock \showarticletitle{{CORD-19:} The Covid-19 Open Research Dataset}.
\newblock \bibinfo{journal}{\emph{CoRR}}  \bibinfo{volume}{abs/2004.10706} (\bibinfo{year}{2020}).
\newblock
\showeprint[arXiv]{2004.10706}
\urldef\tempurl%
\url{https://arxiv.org/abs/2004.10706}
\showURL{%
\tempurl}


\bibitem[Wang et~al\mbox{.}(2022)]%
        {DBLP:conf/sigir/WangMMO22}
\bibfield{author}{\bibinfo{person}{Xiao Wang}, \bibinfo{person}{Sean MacAvaney}, \bibinfo{person}{Craig Macdonald}, {and} \bibinfo{person}{Iadh Ounis}.} \bibinfo{year}{2022}\natexlab{}.
\newblock \showarticletitle{An Inspection of the Reproducibility and Replicability of TCT-ColBERT}. In \bibinfo{booktitle}{\emph{{SIGIR} '22: The 45th International {ACM} {SIGIR} Conference on Research and Development in Information Retrieval, Madrid, Spain, July 11 - 15, 2022}}, \bibfield{editor}{\bibinfo{person}{Enrique Amig{\'{o}}}, \bibinfo{person}{Pablo Castells}, \bibinfo{person}{Julio Gonzalo}, \bibinfo{person}{Ben Carterette}, \bibinfo{person}{J.~Shane Culpepper}, {and} \bibinfo{person}{Gabriella Kazai}} (Eds.). \bibinfo{publisher}{{ACM}}, \bibinfo{pages}{2790--2800}.
\newblock
\urldef\tempurl%
\url{https://doi.org/10.1145/3477495.3531721}
\showDOI{\tempurl}


\bibitem[Yang et~al\mbox{.}(2018)]%
        {DBLP:journals/jdiq/YangFL18}
\bibfield{author}{\bibinfo{person}{Peilin Yang}, \bibinfo{person}{Hui Fang}, {and} \bibinfo{person}{Jimmy Lin}.} \bibinfo{year}{2018}\natexlab{}.
\newblock \showarticletitle{Anserini: Reproducible Ranking Baselines Using Lucene}.
\newblock \bibinfo{journal}{\emph{{ACM} J. Data Inf. Qual.}} \bibinfo{volume}{10}, \bibinfo{number}{4} (\bibinfo{year}{2018}), \bibinfo{pages}{16:1--16:20}.
\newblock
\urldef\tempurl%
\url{https://doi.org/10.1145/3239571}
\showDOI{\tempurl}


\end{thebibliography}

\end{document}